# The correlation dimension of differenced data

Dean Prichard[1]
*Department of Physics,
University of Alaska, Fairbanks, AK 99775*

(Draft: May 25, 1994)

## ABSTRACT

The effect differencing has on the estimated correlation dimension of a nongaussian time series is discussed. Two different methods for generating surrogate data sets are compared. The results suggest that any filtering should be applied to both the raw and surrogate time series, otherwise spurious results can occur.

## 1 Introduction

According to the Takens embedding theorem [1,2], in most cases, both the original series $x_t$ and a differenced series $x_t - x_{t-1}$ should provide a reconstruction of the underlying dynamics. In fact, in cases where there is some drift in the data, one might expect that the differenced time series might even provide a better reconstruction of the dynamics, since differencing can remove nonstationarity in the mean [3]. Another reason for differencing data is that, in many cases, it can be used to remove the linear correlations from the data, and the spurious effects of autocorrelation on dimension estimates (and other nonlinear statistics) are well known [4]. However, differencing does have drawbacks: for example Theiler and Eubank [5] have shown that in many cases pre-whitening of chaotic data sets makes it harder to detect nonlinearity. Also, noise tends to be amplified by differencing.

Many authors have examined differenced time series for evidence of nonlinearity or chaos. For example, Sugihara and May [6] examined the differenced measles and chicken pox data for New York City, as well as a differenced time series of marine phytoplankton populations. Provenzale *et al.* [7] suggested that one should examine both the original and differenced time series, they state that for a stochastic signal the estimated correlation dimension for the differenced series is often much larger than that for the original data, while for a chaotic system the dimensions will be the same. For the examples discussed below, the estimated correlation dimension of a difference stochastic signal turns out to be *smaller* than that for the original data.

---

[1]Current Address: MS-B213, Complex Systems Group, Theoretical Division, Los Alamos National Laboratory, Los Alamos, NM 87545

## 2  Surrogate Data

Theiler *et al.* [8] have advocated a statistical approach for detecting nonlinearity in time series: the approach is to test if the data are consistent with a particular null hypothesis. This is done by generating a ensemble of "surrogate" data sets which reproduce certain linear properties of the original time series but are otherwise stochastic. One then calculates some nonlinear statistic for each of the surrogate sets, and compares the results to those for the same statistic calculated from original data. If the results for the original data are significantly different from the results for the surrogates, the null hypothesis can be rejected. Surrogates to test the null hypothesis that the data is from a linear gaussian stochastic process are easily made by taking the Fourier transform of the data, randomizing the phases and inverting the transform. For data sets which have very nongaussian amplitude distributions, it is clear that the data were not directly generated by a linear gaussian stochastic process, however, such data could be the result of a nonlinear static transform of a linear stochastic process. To test this null hypothesis, one needs surrogate data sets which reproduce the amplitude distribution of the original data, as well as the linear properties. An algorithm to generate such surrogate data sets is provided in Theiler *et al.* [8] (their AAFT method). The idea is to rescale the data so it has a gaussian distribution, make a surrogate data set by the Fourier transform method, and then apply the inverse of the rescaling operation to the surrogate, so the surrogate will have this same distribution as the original data set.

In both of the examples shown below, the Takens estimator of correlation dimension [9,10] is used as the nonlinear statistic

$$D_{\text{Takens}} = \frac{C(r_0)}{\int_0^{r_0} (C(r)/r) dr} \qquad (1)$$

where $r_0$ is a upper cutoff; in this Letter, $r_0$ is set at roughly 1/4 the standard deviation of the time series. $C(r)$ is the correlation integral [11]

$$C(r) = \frac{2}{(N-W)(N-W+1)} \sum_{k=W}^{N} \sum_{i=1}^{N-k} \Theta(r - \|\vec{x}_{i+k} - \vec{x}_i\|) \qquad (2)$$

where $\Theta$ is the Heaviside function, $\|\cdot\|$ is the maximum norm, and $W$ is a constant, the order of a few autocorrelation times, which is used to remove autocorrelative effects [4]. In this Letter, $\vec{x}_t$ is a time delay embedding [12] $\vec{x}_t = (x_t, x_{t-\tau}, \ldots, x_{t-(m-1)\tau})$, where $\tau$ is the time delay and $m$ is the embedding dimension.

## 3  The effect of differencing

In this section it is shown that one should be very careful about differencing (or any other filtering) of time series which have a nongaussian probability distribution. For example, consider a simple AR(1) model: $x_{t+1} = 0.99 x_t + e_t$ which is passed through the nonlinear observation function $y_t = x_t^3$ and then differenced $z_t = y_t - y_{t-1}$. In Fig. 1 the first 8192 points from the $x$, $y$, and $z$ time series are shown. Notice that the $z$ time series has a non-constant variance (hetroscedasticity). This "bursty" behavior is an effect of both the autocorrelation in the data set and the nonlinear measurement function. Below, two possible methods to make surrogates for differenced time series are explored: the first is to make the surrogates directly from the differenced ($z$) time series. The other is to make the surrogates from the original ($y$) time series, and then difference the surrogates. Because of problems associated with the non-constant variance (heteroscedasticity) of the differenced time series, in this case, it is the second method that is preferred (it is assumed that the measurement function is monotonic, for example, it is found that neither of the methods will work if the measurement function is $x^4$).

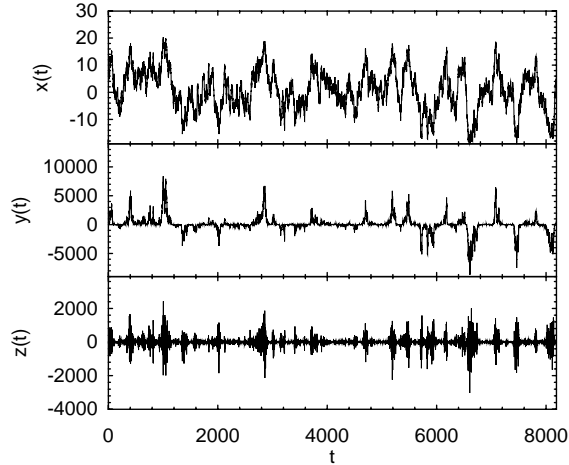

**Fig. 1.** Upper panel: original data set $x_{t+1} = 0.99 x_t + e_t$. Middle panel: time series after transform $y_t = x_t^3$. Lower panel: time series after differencing $z_t = y_t - y_{t-1}$.

In Fig. 2 the differenced time series $z$ is shown, along with two surrogate sets. The surrogate data set shown in the middle panel was made directly from the differenced data, using the amplitude adjusted method (AAFT) of Theiler *et al.* [8] (type I), while that in the lower panel was made from the original data set using the AAFT method, and then differenced (type II). Notice that both the original data and the type II surrogate data seem to come in bursts. That is, the time series is nonstationary in the variance [13] even for very long time periods.

Notice that the type I surrogate does not reproduce bursty behavior of the original signal, and it is known that this kind of behavior can lead to spurious estimates of the correlation dimension [14]. The inability of the type I surrogate to mimic the properties of the differenced data set is due to the fact that the differencing operation and the transforming operation do not commute. That is, if one takes a realization of a linear gaussian process, applies the differencing operator to it, and then does a nonlinear transform of the data, what one gets is a nonlinear static transform of a linear process. However, if one takes a realization of linear gaussian process, applies a nonlinear static transform, and then does the differencing, the result is *not* a static nonlinear transform of a linear stochastic process. This suggests that in many cases (especially when the measurement function is unknown) one should make the surrogates from the original time series and then apply any "filtering" to both the original and surrogate data sets.

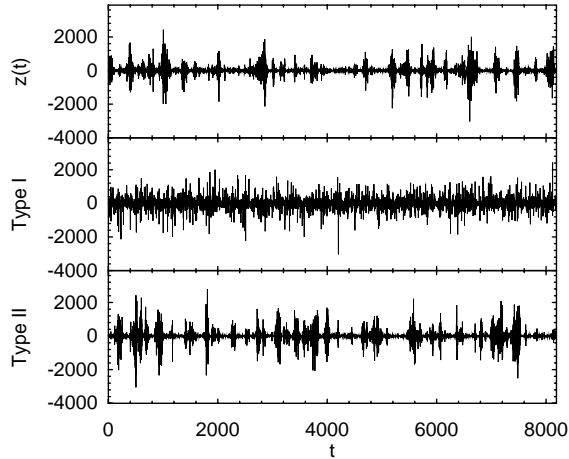

**Fig. 2.** Upper panel: $z$ time series. Middle panel: Type I surrogate data set. Lower panel: Type II surrogate data set.

## 3.1 Correlation dimension

By looking at Fig. 2, one can see that the type I surrogates do not reproduce the properties of the differenced data. In this section, it is shown that if one does not use surrogate data sets which reproduce this hetroscedasticity, one might identify a linear stochastic process as a nonlinear one. In Fig. 3a the Takens estimator with a upper cutoff equal to 1/4 the standard deviation of the $z$ time series is shown, using a time delay $\tau = 1$, $W = 20$ and embedding dimensions $m = 1, \ldots, 10$ for the original data (solid), and for 39 type I surrogates (dots) Fig. 3b shows the same statistic, but this time using the type II surrogates (with 39 surrogate data sets if the value of the statistic for the original data set falls outside the distribution of the statistic for the surrogates, one can reject the null hypothesis that the data is linear at the 95% level). The underlying process is linear, therefore, there should be no substantial differences between the value of the Takens estimator for the original data and that for the surrogate data sets. Since the value of the Takens estimator for the type I surrogate is in all cases greater than that for the original data, one can see that it is the type II surrogates which are preferred. This demonstrates why it is important to make surrogate data sets which reproduce the non-constant variance of the original data, otherwise, spurious detection of nonlinearity can occur.

## 4 Application to real data

Recently there has been much interest in attempting to use methods of nonlinear time series analysis to characterize the global dynamics of the magnetosphere, based on the AE and AL indices (see Takalo *et al.* [15], Prichard and Price [16] and references within). Many of the more recent studies show there is no evidence that the magnetosphere can be described as a low dimensional autonomous nonlinear system, in contrast with earlier results (see Roberts *et al.* [17] and references within). However, recently some authors have claimed to find evidence for low dimensions based on examining the differenced AL time series [18,19]. One reason for differencing is that the AL index is nonstationary for time series shorter than a few months [20]. Also, differencing the AL index removes most of the autocorrelation.

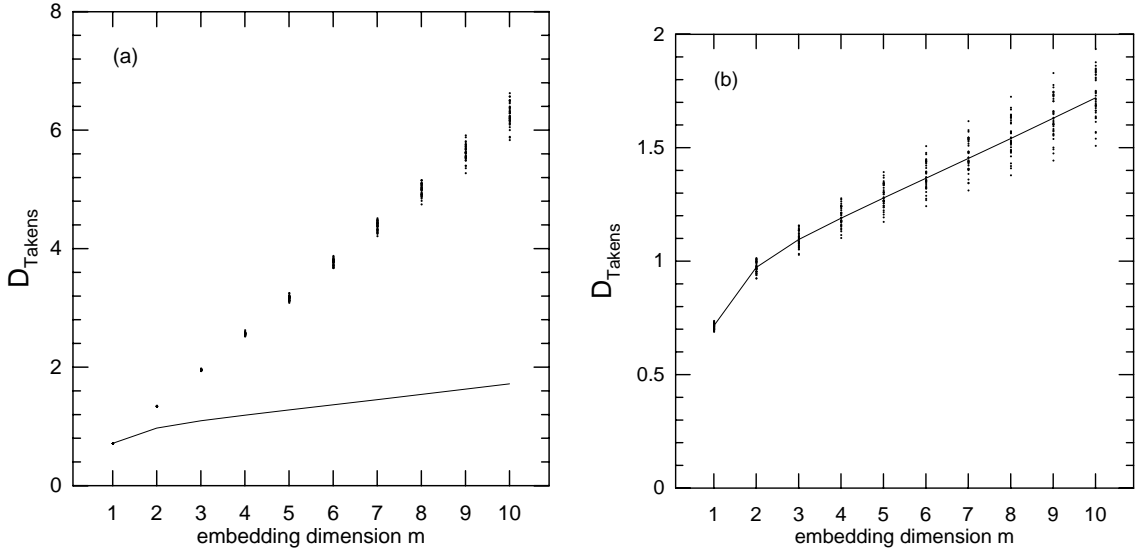

**Fig. 3.** (a) Takens dimension estimator as a function of embedding dimension for the $z$ time series (solid) and 39 type I surrogates. (b) Same but for the type II surrogates.

This is important, as it has been previously shown [21] that many of the previous estimates of dimensions for the AE and AL indices were most likely caused by the autocorrelation of the data [4], and are not the result of low dimensional dynamics.

In Fig. 4 the differenced AL index for the first 8192 points of January 1983 is shown along with two surrogate sets. The surrogate data set shown in the middle panel was made directly from the differenced data, using the AAFT method of Theiler *et al.* [8] (type I), while that in the lower panel was made from the original data set using the AAFT method, and then differenced (type II). Notice that both the original data and the type II surrogate seem to come in bursts.

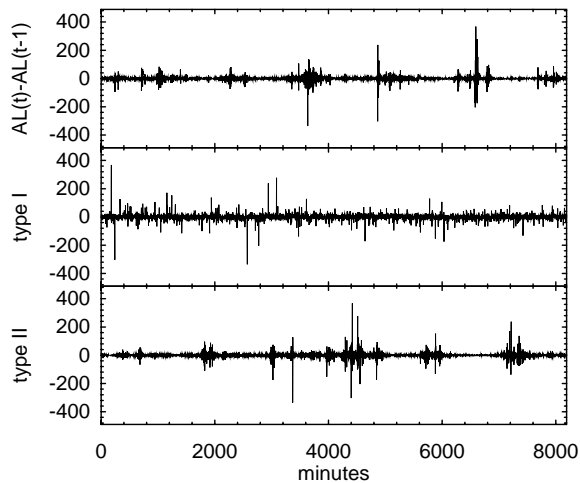

**Fig. 4.** Upper panel: first difference of AL index for the first 8192 minutes of January 1983. Middle panel: Type I surrogate data set. Lower panel: Type II surrogate data set.

## 4.1 Correlation dimension

Previous studies have found correlation dimensions between 2.5 and 3.0 [18, 19] for the differenced AL index. In this section, 32768 points of the differenced AL index from the month of January 1983 are analyzed. Following the previous authors, embedding dimensions $m = 1, \ldots, 10$ and a time delay of $\tau = W = 5$ are used. In Fig. 5a the Takens estimator with $r_0 = 5$ nT (which is where there is a plateau in the slope of the $\log C(r)$ versus $\log r$ curves) is shown as a function of embedding dimension for the differenced AL data (solid line) and for the 39 type I surrogates (dots). There is a clear distinction between the dimensions for the type I surrogates and that for the original data set. This is the same result obtained by Sharma *et al.* [19], that is, randomizing the phases destroys the convergence of the dimension. In Fig. 5b the results using the type II surrogates, which reproduce the non-constant variance, are shown. There is no difference between the estimated dimensions of the type II surrogates and those for the original data set.

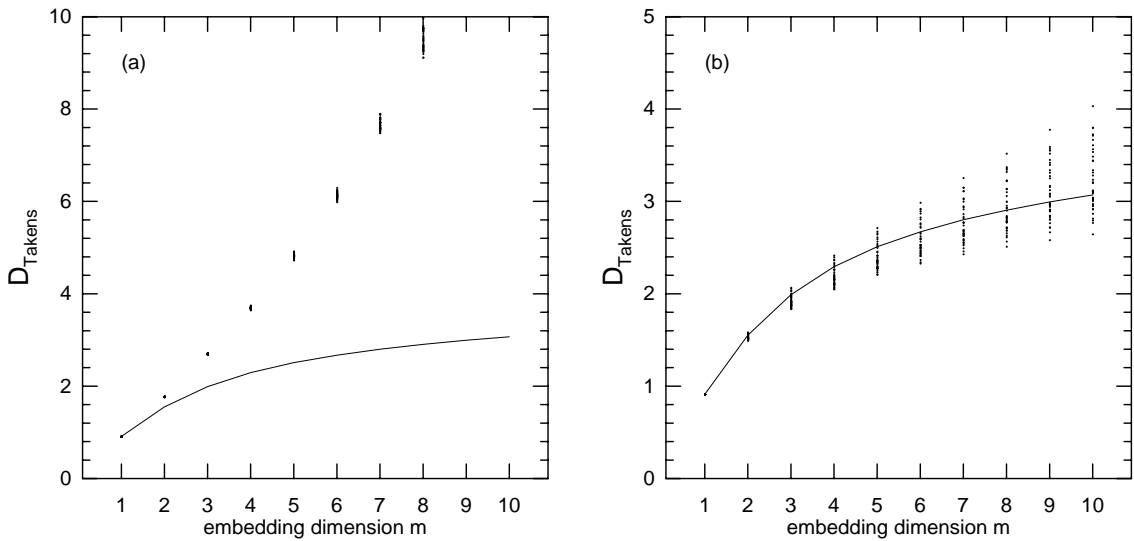

Fig. 5. (a) Takens dimension estimator versus embedding dimension for the differenced AL data (solid line), and the type I surrogates (dots). (b) Same for the type II surrogates.

This suggests that, for this interval, the AL index is consistent with a static nonlinear transform of a linear stochastic process. The previous estimates of low dimensions for the differenced AL data were most likely an effect of this non-constant variance. As noted by Grassberger *et al.* [14] this sort of "bursty" behavior leads to very non-homogeneous distribution of points in the reconstructed state space, where many pairs of points used in the correlation integral come from a region of state space which is basically a single point, and this trivially leads to a severe underestimate of the dimension.

## 5 Conclusions

It has been shown that the amplitude adjusting method described in Theiler *et al.* [8], and linear filtering do not commute. Because of this, it is suggested that one should not filter nongaussian data before making surrogate data sets, but instead make the surrogates from the original data and apply the same filtering to both the original and surrogate data sets. By applying both of these methods to generate surrogate data to a time series of the differenced AL index, it is seen that the previous estimates of low dimensions for this data set were most likely an artifact of the non-constant variance of the data, and not the results of low dimensional magnetospheric dynamics.

**Acknowledgments**

I thank James Theiler and Channon Price for many useful discussions. This work was partially supported by NSF grants ATM-9213522 and ATM-9311522.